\documentclass[nofootinbib,showpacs,preprintnumbers,amsmath,amssymb]{revtex4}
\usepackage{epsfig}

\begin{document}

\title{Non-singlet structure functions: Combining the leading
logarithms resummation at small -$x$ with DGLAP}

\vspace*{0.3 cm}

\author{B.I.~Ermolaev}
\affiliation{Ioffe Physico-Technical Institute, 194021
 St.Petersburg, Russia}
\author{M.~Greco}
\affiliation{Department of Physics and INFN, University Rome III,
Rome, Italy}
\author{S.I.~Troyan}
\affiliation{St.Petersburg Institute of Nuclear Physics,
188300 Gatchina, Russia}

\begin{abstract}
The explicit expressions for the non-singlet DIS structure
functions $F_{1}$ and $g_1$, obtained at small $x$ by resumming
the leading logarithmic contributions to all orders, are discussed
and compared in detail with the DGLAP evolution for different
values of $x$ and $Q^2$. The role played by the DGLAP inputs for
the initial parton densities on the small-$x$ behavior of the
non-singlet structure functions is discussed. It is shown that the
singular factors included into the fits ensure the Regge
asymptotics of the non-singlet structure functions and mimic the
impact of the total resummation of the leading logarithms found
explicitly in our approach. Explicit expressions are presented
which implement the NLO DGLAP contributions with our small -$x$
results.
\end{abstract}


\maketitle

\section{Introduction}

The non-singlet components of the structure function $F_1$
$(\equiv f^{NS})$ and of the spin structure function $g_1$
$(\equiv g_1^{NS})$ have been investigated in great detail in deep
inelastic scattering (DIS) experiments. The standard theoretical
framework is provided by DGLAP\cite{dglap}. In this approach, any
of $ f^{NS}(x, Q^2), g_1^{NS}(x, Q^2)$ can be represented as a
convolution of the coefficient functions and the evolved quark
distributions calculated in LO\cite{dglap}, NLO\cite{fp} and
NNLO\cite{nv}. Combining these results with appropriate fits for
the initial quark distributions, provides a good agreement with
the available experimental data (see e.q.\cite{a} and the recent
paper \cite{v})

On the other hand, the DGLAP evolution equations were originally
applied in a range of rather large $x$ values, where higher-loop
contributions to the coefficient functions and the anomalous
dimensions are small. According to the results of
Refs~\cite{ber,egt1}, such corrections are becoming more and more
essential when $x$ is decreasing and DGLAP should  not work so
well at $x\ll 1$. Nevertheless, it turns out that DGLAP
predictions are in a good agreement with available experimental
data, leading to the conclusion that the impact of the
higher-order corrections is negligibly small for the available
values of $x$ and those corrections may be relevant at
asymptotically small $x$ only.

In the present paper we use our previous results \cite{egt1,egt2}
to show  that the impact of the high-order corrections on the
$Q^2$ and $x$ -evolutions of the non-singlet structure functions
is indeed quite sizable at available values of $x$. We also show
that the reason for the success of phenomenological analysis of
the data based on DGLAP at $x < 10^{-2}$ is related to the sharp
$x$ -dependence assumed for the initial parton densities, which is
able to mimic the role of high-order corrections.

The paper is organized as follows: In Sect.~2 we introduce our
notations and discuss the difference of our approach with DGLAP.
In Sect.~3 the DGLAP and our predictions for the small-$x$
asymptotics are explicitly compared. It allows us to clarify the
role played by the singular terms in $x$ in the DGLAP inputs for the
initial parton densities. In Sect.~4 we give a detailed numerical
comparison of our results with DGLAP, and discuss also the role played by
the regular part of the initial parton densities.
 As our approach deals with small $x$ only, in Sect.~5 we suggest
 a possible method to combine DGLAP with
our approach in order to obtain equally correct expressions for
both large and small values of $x$. Finally, Sect.~5 contains our
conclusions.

\section{Comparison between DGLAP and our approach}

As the DGLAP -expressions for the non-singlet structure functions
are well-known, we discuss them briefly only. In this approach,
both $ f^{NS}(x, Q^2)$ and $g_1^{NS}(x, Q^2)$ $(\equiv
f_{DGLAP}(x, Q^2))$ can be represented as a convolution
\begin{equation}
\label{fdglap} f_{DGLAP}(x, Q^2) = \int_x^1
\frac{dy}{y}C(x/y)\Delta q(y, Q^2)
\end{equation}
of the coefficient functions $C(x)$ and the evolved quark
distributions $\Delta q(x, Q^2)$. Similarly, $ d \Delta q(x, Q^2)/
\ln(Q^2)$ can be expressed through the convolution of the
splitting functions and the initial quark densities $\delta q(x
\approx 1, Q^2 \approx \mu^2)$ where $\mu^2$ is the starting point
of the $Q^2$ -evolution. It is convenient to represent
$f_{DGLAP}(x, Q^2)$ in the integral form, using the Mellin
transform:

\begin{equation}
\label{fdglapmellin} f_{DGLAP}(x, Q^2) = (e^2_q/2) \int_{-\imath
\infty}^{\imath \infty} \frac{d \omega}{2\imath \pi}(1/x)^{\omega}
C(\omega) \delta q(\omega) \exp \Big[\gamma(\omega)
\int_{\mu^2}^{Q^2} \frac{d k^2_{\perp}}{k^2_{\perp}}
\alpha_s(k^2_{\perp})\Big]
\end{equation}
where $C(\omega)$ are the non-singlet coefficient functions,
$\gamma(\omega)$ the non-singlet anomalous dimensions and $\delta
q(\omega)$ the Mellin transforms of the initial non-singlet quark
densities. In the $x$- space, the standard DGLAP inputs for $\delta
q(x)$ for the non-singlet parton densities (see e.g.
Refs.~\cite{a, v}) include terms which are singular when $x \to 0$,
as well as regular ones for all $x$. For example, the fit A of
Ref.\cite{a} is chosen as follows:

\begin{eqnarray}
\label{fita} \delta q(x) = N \eta x^{- \alpha}(1 -x)^{\beta}(1 +
\gamma x^{\delta}) = &&N \eta
 x^{-\alpha}\phi(x)~, \\ \nonumber
&&\phi(x) \equiv (1 -x)^{\beta}(1 + \gamma x^{\delta})~,
\end{eqnarray}
with $N,~\eta$ being normalization factors, $\alpha = 0.576$, $\beta =
2.67$, $\gamma = 34.36$ and $\delta = 0.75$. As the first term $N
\eta x^{-\alpha}$ in the rhs of Eq.~(\ref{fita}) is singular when
$x \to 0$ whereas the second one, $\phi (x)$ is regular, we will
address them as the singular and regular parts of the fit
respectively. Obviously, in the $\omega$ -space Eq.~(\ref{fita})
is a sum of pole contributions:
\begin{equation}
\label{fitaomega} \delta q(\omega) = N \eta \Big[ (\omega -
\alpha)^{-1} + \sum_{k = 1}^{\infty} m_k \Big((\omega + k -
\alpha)^{-1} + \gamma (\omega + k +1 - \alpha)^{-1}\Big)\Big]~,
\end{equation}
with $m_k = \beta (\beta - 1)..(\beta - k + 1)/ k!$, so that the
first term in Eq.~(\ref{fitaomega}) (the leading pole) corresponds
to the singular term $x^{-\alpha}$ of Eq.~(\ref{fita}) and the
second term, i.e. the sum of the poles, corresponds to the
interference between the singular and regular terms. In contrast
to the leading pole position $\omega = \alpha$, all other poles in
Eq.~(\ref{fitaomega}) occur at negative values, because $k - \alpha >
0$. As usual, when $\omega$ is integer:~$\omega = n = 1,2,..$, the
integrand
of Eq.~(\ref{fdglapmellin}) gives the $n$-th moment of $f$,
$(\equiv \Gamma_n(Q^2))$. The standard DGLAP -technology of
calculating $f(x, Q^2)$ corresponds to first calculating
$\Gamma_n(Q^2)$ and then reconstructing  $f(x, Q^2)$ from the
moments, choosing appropriate forms for $\Delta q$. Presently
$C(\omega)$ for the non-singlet structure functions are known with
two-loop \cite{fp} and three-loop\cite{nv} accuracy.

In order to make the all-order resummation of the
double-logarithmic contributions to  $ f^{NS}(x, Q^2)$ and
$g_1^{NS}(x, Q^2)$, i.e. the contributions $\sim (\alpha_s/
\omega^2)^k$ to $C(\omega)$ and $\gamma(\omega)$ in
Eq.~(\ref{fdglapmellin}), an alternative approach was used in
Refs.~\cite{ber}, by introducing and solving infrared evolution
equations with fixed $\alpha_s$. This approach was improved in
Refs.~\cite{egt1}, where single-logarithmic contributions were
also accounted for and the QCD coupling was running in the Feynman
graphs contributing to the non-singlet structure functions. In
contrast to the DGLAP parametrization $\alpha_s =
\alpha_s(k^2_{\perp})$, we used in Refs.~\cite{egt1} another
parametrization where the argument of $\alpha_s$ in the quark
ladders is given by the time-like virtualities of the intermediate
gluons. Arguments in favor of such a parametrization were given in
Ref.~\cite{egt2}. In particular, it was shown that it coincides
with the DGLAP -parametrization when $x \sim 1$, but it differs
sensibly when $x \ll 1$. Instead of Eq.~(\ref{fdglapmellin}),
Refs.~\cite{egt1} suggest the following formulae for the
non-singlet structure functions \footnote{Integration over
$\omega$ in Eq.~(\ref{fnsint}) was performed in
Refs.~\cite{sman},\cite{lubl}, however with simplified assumptions
for $\alpha_s$.}:

\begin{equation}
\label{fnsint}
f^{NS}(x, Q^2) = (e^2_q/2) \int_{-\imath \infty}^{\imath \infty}
\frac{d \omega}{2\pi\imath}(1/x)^{\omega}
C_{NS}^{(+)}(\omega)
\delta q(\omega)
\exp\big( H_{NS}^{(+)}(\omega) y\big)
\end{equation}

\begin{equation}
\label{gnsint}
g_1^{NS}(x, Q^2) = (e^2_q/2) \int_{-\imath \infty}^{\imath \infty}
\frac{d \omega}{2\pi\imath }(1/x)^{\omega}
C_{NS}^{(-)}(\omega)
\delta q(\omega)
\exp\big( H_{NS}^{(-)}(\omega) y\big)~,
\end{equation}
with $y = \ln(Q^2/ \mu^2)$ so that $\mu^2$ is the starting point of the
$Q^2$ -evolution.
The new coefficient functions  $C_{NS}^{(\pm)}$ are expressed in terms of
new anomalous dimensions $H_{NS}^{(\pm)}$:
\begin{equation}
\label{cns}
C_{NS}^{(\pm)} =\frac{\omega}{\omega - H_{NS}^{(\pm)}(\omega)}
\end{equation}
and the new anomalous dimensions  $H_{NS}^{(\pm)}(\omega)$ which
account for
the resummation of the double- and single- logarithmic
contributions are

\begin{equation}
\label{hns}
H_{NS}^{(\pm)} = (1/2) \Big[\omega -
\sqrt{\omega^2 - B^{\pm}(\omega)} \Big]
\end{equation}
where
\begin{equation}
\label{b} B^{(\pm)}(\omega) = (4\pi C_F (1 +  \omega/2) A(\omega)
+ D^{(\pm)}(\omega))/ (2 \pi^2)~.
\end{equation}
Finally $ D^{(\pm)}(\omega)$ and $A(\omega)$ in Eq.~(\ref{b}) are
expressed in
terms of  $\rho = \ln(1/x)$, $b = (33 - 2n_f)/12\pi$ and the color factors
 $C_F = 4/3$, $N = 3$:

\begin{equation}
\label{d}
D^{(\pm)}(\omega) = \frac{2C_F}{b^2 N} \int_0^{\infty} d \eta
e^{-\omega \eta} \ln \big( \frac{\rho + \eta}{\eta}\big)
\Big[ \frac{\rho + \eta}{(\rho + \eta)^2 + \pi^2} \mp \frac{1}{\eta}\Big] ~,
\end{equation}

\begin{equation}
\label{a}
A(\omega) =
\frac{1}{b} \Big[\frac{\eta}{\eta^2 + \pi^2} - \int_0^{\infty}
\frac{d \rho e^{-\omega \rho}}{(\rho + \eta)^2 + \pi^2} \Big]
\end{equation}

$A(\omega)$ is the Mellin transform of $\alpha_s$ with the
time-like argument $k^2$,
 $\alpha_s(k^2) = 1/(b \ln(-k^2/\Lambda^2))$. The
comparison of  Eq.~(\ref{fdglapmellin}) to Eqs.~(\ref{fnsint},
\ref{gnsint}) clearly
shows the difference between DGLAP and our approach:

First, our approach accounts for the double- and single- logarithmic
contributions to all orders in the QCD coupling whereas in the
DGLAP -expressions the coefficient functions and
the anomalous dimensions are obtained up to fixed-order accuracy (two and
three loops).

Then a second difference is related to the treatment of
$\alpha_s$. In DGLAP, the argument of  $\alpha_s$ in any Feynman
graph is $k^2_{\perp}$, with $k$ being the momentum of the ladder
quarks, and therefore  $\alpha_s$ is not involved in the Mellin
transform. This leads to the $Q^2$ dependence of the non-singlet
structure functions with the exponent $\sim \ln \ln Q^2$. In our
approach  $\alpha_s$ depends on time-like virtualities of the
ladder gluons and therefore, it is involved in the Mellin
transform.

The inclusion of $\alpha_s$ in the Mellin transform explicitly
shows that the factorization between the longitudinal and
transverse spaces used in DGLAP is not valid for small $x$.
However, as shown in Ref.~\cite{egt2}, our approach and DGLAP
converge when $x \sim 1$ where the factorization does hold. The
parametrization of $\alpha_s(k^2)$ in the form of Eq.~(\ref{a})
has recently been used in Refs.~\cite{kotl} for calculating the
double- and single- logarithmic corrections to the Bjorken sum
rules.

\section{The role of the singular factor for $\delta q$ in DGLAP fits}

In the first place let us study the small-$x$ asymptotics of the
non-singlet structure functions. When $x \to 0$, one can use the
saddle point method in order to estimate the integrals in
Eqs.~(\ref{fnsint},\ref{gnsint}) and derive much simpler
expressions for the non-singlet structure functions. When the
initial parton densities $\delta q(x)$ are regular in $x$ (e.g.
given by $\phi(x)$ only in Eq.~(\ref{fita})), they do not contribute to
the saddle point position, which is:

\begin{equation}
\label{saddle} \omega_0^{(\pm)} = \sqrt{B^{(\pm)}} \Big[ 1 + (1 -
q^{(\pm)})^2 (y/2 + 1/\sqrt{B^{(\pm)}})^2/(2\ln^2 \xi) \Big]~,
\end{equation}
with
\begin{equation}
\label{q}
 q^{(\pm)} = d \sqrt{B^{(\pm)}}/ d\omega |_{\omega = \omega_0^{(\pm)}}
\end{equation}
and $y = \ln(Q^2/\mu^2)$, $\xi = Q^2/(x^2 \mu^2)$.

It immediately leads to the Regge asymptotics for the
non-singlets:

\begin{equation}
\label{as} f^{NS} \sim e_q^2 \delta q(\omega_0^{(+)})
\Pi^{(+)}_{NS} \xi^{\omega_0^{(+)}/2},~~~ g_1^{NS} \sim e_q^2
\delta q(\omega_0^{(-)}) \Pi^{(-)}_{NS} \xi^{\omega_0^{(-)}/2}~,
\end{equation}
with
\begin{equation}
\label{pi} \Pi^{(\pm)}_{NS} = \frac{\Big[2(1 -
q^{(\pm)})\sqrt{B^{(\pm)}}\Big]^{1/2} \big(y/2 +
1/\sqrt{B^{(\pm)}} \big)}{\pi^{1/2}\ln^{3/2}\xi}~.
\end{equation}

For $Q^2 \ll 150~$GeV$^2$, $y \ll 2/\sqrt{B^{(\pm)}}$, so
$\omega_0^{(\pm)}$ do not depend on $y$.
 Therefore Eq.~(\ref{as}) can be rewritten as:

\begin{equation}
\label{asympt}
f^{NS}(x,Q^2)
\sim e_q^2 \delta q(\omega_0^{(+)})
c^{(+)}  T^{(+)}(\xi),
~~~g_1^{NS}(x,Q^2) \sim e_q^2 \delta q(\omega_0^{(-)})
c^{(-)} T^{(-)}(\xi)
\end{equation}
with
\begin{equation}
\label{t}
 T^{(\pm)}(\xi) = \xi^{\omega_0^{(\pm)}/2}/\ln^{3/2}\xi
\end{equation}
and the intercepts $\omega_0^{(+)} = 0.38$ and
$\omega_0^{(-)}= 0.43$ (see \cite{egt1} for details). The factors
$c^{(\pm)}$,
\begin{equation}
\label{c}
 c^{(\pm)} = \Big[2(1 - q)/(\pi \sqrt{B^{(\pm)}}) \Big]^{1/2}
\end{equation}
do not depend on $y$. Eq.~(\ref{asympt}) predicts the
asymptotic scaling for the non-singlet structure functions: in the region
 $Q^2\ll~150~$GeV$^2$, with $T^{(\pm)}$ depending on one argument $\xi$
instead of $x$ and $Q^2$ independently. In the opposite case when
 $Q^2 > ~150~$GeV$^2$, the scaling is valid for $f^{NS}/y$ and
$g_1^{NS}/y$. The small-$x$ dependence of Eq.~(\ref{asympt}) is
confirmed by various analyses of the experimental data in
Refs.~\cite{kat}, while
the $Q^2$- dependence of these formulae has not been checked yet.

On the other hand, under the same hypothesis that $\delta q(x)$ is regular
(e.g. given again
by $\phi(x)$ in Eq.~(\ref{fita})), the small-$x$ asymptotics of
$f_{DGLAP}$ is well-known (we drop the inessential pre-factor
here):
\begin{equation}
\label{asdglapreg} f_{DGLAP}^{NS} \sim \exp \sqrt{\big((2 C_F)/\pi
b \big)\ln(1/x) \Big((\ln(Q^2/\Lambda^2))/
(\ln(\mu^2/\Lambda^2))\Big)} ~.
\end{equation}
Obviously, this behavior is different and less steep than the
Regge asymptotics of Eq.~(\ref{asympt}). However when the singular
part in $\delta q$ is also included in the DGLAP input of
Eq.~(\ref{fita}), the DGLAP asymptotics for the non-singlet
structure functions is controlled by the leading singularity
$\omega = \alpha = 0.576$ of the fit in the integrand in
Eq.~(\ref{fitaomega}) so that the asymptotics of $f_{DGLAP}(x,
Q^2)$ is also of the Regge type:
\begin{equation}\label{asdglap}
f_{DGLAP}^{NS} \sim (e^2_q/2)
C(\alpha)(1/x)^{\alpha}\Big((\ln(Q^2/\Lambda^2))/
(\ln(\mu^2/\Lambda^2))\Big)^{\gamma(\alpha)/b}~,
\end{equation}
with $b = (33 - 2n_f)/12\pi$. The comparison of Eq.~(\ref{asympt})
and Eq.~(\ref{asdglap}) shows that both DGLAP and our approach
lead to the Regge asymptotic behavior in $x$ for the non-singlet
structure functions. In particular, Eq.~(\ref{fita}) makes the
DGLAP prediction be more singular in $x$ than ours, Also they
differ as far as the predictions for the $Q^2$ -behavior are
concerned. However, it is important to stress that our intercepts
$\omega_0^{\pm}$ are naturally obtained by the total resummation
of the leading logarithmic contributions, without assuming any
singular input for $\delta q$, whereas the DGLAP intercept
$\alpha$ in Eq.~(\ref{asdglap}) is generated by the
phenomenological factor $x^{-0.57}$ of Eq.~(\ref{fita}), which
makes the structure functions grow when $x$ decreases, and mimics
in fact the total resummation\footnote{We remind that our
estimates for the intercepts $\omega_0^{\pm}$ were confirmed by
several analyses\cite{kat} of the experimental data}. In other
words, the role of the higher-loop radiative corrections on the
small-$x$ behavior of the non-singlets is, actually, incorporated
into DGLAP phenomenologically, through the initial parton
densities fits. It also means that the singular factors can be
dropped from such fits
 when the coefficient functions account for the total resummation
of the leading logarithms and therefore regular functions of $x$ only
can be chosen  for the initial densities $\delta q$.

\section{The role of the regular part in the initial parton densities}

As discussed in Sect. II, the regular part in the initial quark
distribution in the DGLAP approach, e.g.
$\phi(x) = N (1
-x)^{\beta}(1 + \gamma
x^{\delta})$ in Eq.~(\ref{fita}), corresponds to
the sum of pole contributions $\phi(\omega)$:
\begin{equation}\label{regomega}
\phi(\omega) \equiv N \sum_{k = 1}^{\infty} m_k \Big((\omega + k
)^{-1} + \gamma (\omega + k +1)^{-1}\Big)\Big]~.
\end{equation}
We have introduced here the normalization factor $N$.
Obviously, $\phi \to N$ when $x \to 0$. An input $\delta q(x) =
N$ means that the shape of the initial quark distribution does not
depend on $x$. In order to study the influence of the initial parton
densities
in more detail, we compare
numerically our results (\ref{gnsint}) for two
situations: (a) a regular fit $\delta q(x) = \phi(x)$ is used,
(b) $\delta q(x) = N$. We define $R_1(x)$ as follows:

\begin{equation}
\label{r1}
 R_1(x) = g_1^{NS}\big[ \delta q(x) = \phi(x)\big]/g_1^{NS}\big[ \delta q(x) = N \big]
\end{equation}
at $Q^2 = 10~$GeV$^2$. For all numerical studies in the present
section we choose the starting point of the $Q$ -evolution $\mu =
1.5~$GeV and assume $\Lambda = 0.15$~GeV and $n_f = 4$. The
results for $R_1$ are plotted in Fig.~1 (curve~1). It shows that
one can approximate $\phi(x)$ by a constant distribution for $x <
0.05$ whereas at larger values of $x$ the $x$ -dependence of the
initial quark distributions is essential.

The comparison of our results (\ref{gnsint}) with
DGLAP when the regular fit $\phi(x)$ is used is also
shown in Fig.~1 (curve~2) for the ratio $R_2$ defined as
\begin{equation}
\label{r2}
 R_2(x) =
 g_1^{NS}\big[ \delta q(x) = \phi(x)\big]/g^{NS}_{1~DGLAP}\big[ \delta q(x) = \phi(x)
 \big]
\end{equation}
where we choose again $Q^2 = 10~$GeV$^2$. This shows that when the
input for $\delta q$ does not have a singular term, the difference
between our results and DGLAP grows fast for $x < 0.05$. On the
contrary, when the singular term is included into the DGLAP input, the
DGLAP results grow faster than ours. This is shown in Fig.~1
(curve~3) for the ratio $R_3$:

\begin{equation}
\label{r3}
 R_3(x) =
 g_1^{NS}\big[ \delta q(x) = \phi(x)\big]/
 g^{NS}_{1~DGLAP}\big[ \delta q(x) = x^{-\alpha} \phi(x) \big]~,
\end{equation}
with $Q^2 = 10~$GeV$^2$. A fast growth of $f^{NS}_{DGLAP}\big[
\delta q(x) = x^{-\alpha} \phi(x) \big]$ for small $x$ also contradicts
the analyses of experimental data obtained in
Refs.~\cite{kat}.

Our results (\ref{fnsint},\ref{gnsint}) differ slightly from DGLAP
also for the $Q^2$ -evolution, depending on x.  This is shown in
Fig.~2 where the ratio $R(Q^2)$:

\begin{equation}
\label{rq}
 R(Q^2) = g_1^{NS}\big[ \delta q(x) = \phi(x)\big]
 /g^{NS}_{1~DGLAP}\big[ \delta q(x) = \phi(x)\big]
\end{equation}
is plotted for $x =0.01$ (curve~1),  $x =0.001$ (curve~2) and for $x
=0.0001$
(curve~3). Fig.~2 also suggests that the difference between our and
DGLAP predictions for the $Q^2$ -evolution slowly grows with
increasing $Q^2$.

\section{Implementation of DGLAP with our higher-loop contributions}

We have shown that Eqs.~(\ref{fnsint},\ref{gnsint}) account for
the resummation of the double- and single logarithmic
contributions to the non-singlet anomalous dimensions and the
coefficient functions that are leading when $x$ is small. However,
the method we have used does not allow us to account for all other
contributions which can be neglected for $x$ small but become
quite important when $x$ is not far from 1. On the other hand,
such contributions are naturally included in DGLAP, where the
non-singlet coefficient function $C_{DGLAP}$ and anomalous
dimension $\gamma_{DGLAP}$ are known with the two-loop
accuracy\footnote{See for details Ref.~\cite{grsv} and the
review\cite{neerv}}:

\begin{eqnarray}
\label{formdglap} &&C_{DGLAP} = 1 + \frac{\alpha_s(Q^2)}{2\pi}
C^{(1)}~, \\ \nonumber &&\gamma_{DGLAP} =
\frac{\alpha_s(Q^2)}{4\pi}\gamma^{(0)} +
\Big(\frac{\alpha_s(Q^2)}{4\pi}\Big)^2 \gamma^{(1)}~.
\end{eqnarray}

Therefore, we can borrow from the DGLAP formulae the contributions
which are missing in
Eqs.~(\ref{fnsint},\ref{gnsint}) by adding
$C_{DGLAP}$ and $\gamma_{DGLAP}$ to the coefficient functions and
anomalous dimensions of Eqs.~(\ref{fnsint},\ref{gnsint}). Of course, both
$C_{DGLAP}$ and $\gamma_{DGLAP}$ contain also terms already
included in Eqs.~(\ref{fnsint},\ref{gnsint}), so in order to avoid a
double counting these terms  should be
extracted from $C_{NS}^{(\pm)},~H_{NS}^{(\pm)}$.

In order to do so, let us consider the region of $x \sim 1$ where the
effective values of $\omega$ in Eqs.~(\ref{fnsint},\ref{gnsint})
are large. In this region we can expand $H_{NS}^{(\pm)}$ and
$C_{NS}^{(\pm)}$
into a series in $1/\omega$. Retaining the first two terms in each series
and
expressing them through $A$ and $D^{(\pm)}$,
 we arrive at $C_{NS}^{(\pm)} = \widetilde{C}_{NS}^{(\pm)} +
O(\alpha_s^2)$, $H_{NS}^{(\pm)} = \widetilde{H}_{NS}^{(\pm)} +
O(\alpha_s^3)$ with

\begin{eqnarray}
\label{series} &&\widetilde{C}_{NS}^{(\pm)} = 1 +
\frac{A(\omega)C_F}{2\pi}\big[ 1/\omega^2 + 1/2\omega\big]~,
\\ \nonumber
&&\widetilde{H}_{NS}^{(\pm)} = \frac{A(\omega)C_F}{4\pi} \big[
2/\omega + 1 \big] +\Big(\frac{A(\omega)C_F}{4\pi}\Big)^2
(1/\omega)\big[ 2/\omega + 1 \big]^2  + D^{(\pm)}(\omega)[1/\omega
+ 1/2]~.
\end{eqnarray}
Now let us define the new coefficient functions
$\hat{C}_{NS}^{(\pm)}$ and new anomalous dimensions
$\hat{C}_{NS}^{(\pm)}$ as follows:

\begin{eqnarray}
\label{newch} &&\hat{H}_{NS}^{(\pm)} = \Big[H_{NS}^{(\pm)} -
\widetilde{H}_{NS}^{(\pm)}\Big] +
\frac{A(\omega)}{4\pi}\gamma^{(0)} +
\Big(\frac{A(\omega)}{4\pi}\Big)^2 \gamma^{(1)}~, \\
\nonumber &&\hat{C}_{NS}^{(\pm)} = \Big[C_{NS}^{(\pm)} -
\widetilde{C}_{NS}^{(\pm)} \Big] +
 1 + \frac{A(\omega)}{2\pi} C^{(1)}~.
\end{eqnarray}

These new, "implemented" coefficient functions
and anomalous dimensions
of Eq.~(\ref{newch}) include both the total resummation of the leading
contributions and the DGLAP expressions in which $\alpha_s(Q^2)$
is replaced by $A(\omega)$.
As shown in Ref.~\cite{egt2}, this parametrization
differs from the DGLAP one, $\alpha_s =
\alpha_s(k^2_{\perp})$, at small $x$, though both parameterizations
coincide when $x$ is large. The main point is that the factorization of
the phase space into transverse and longitudinal spaces used
in DGLAP is a good approximation for large $x$ only.
The implemented coefficient functions $\hat{C}_{NS}^{(\pm)}$ and
anomalous dimensions $\hat{H}_{NS}^{(\pm)}$ of Eq.~(\ref{newch})
can be used either in the standard DGLAP way by letting $\omega$
be integer, i.e. $\omega = n$, with $n = 1,2,...$ and thereby obtaining
the moments of the non-singlet structure functions, or alternatively by
using
$\hat{C}_{NS}^{(\pm)}, \hat{H}_{NS}^{(\pm)}$ in
Eqs.~(\ref{fnsint},\ref{gnsint}) where $\omega$ is complex. In the
latter case, $C_{DGLAP}$ and $\gamma_{DGLAP}$ should first be
continued analytically from the integer values $\omega = n$.
We have verified that the new formulae in  Eq.~(\ref{newch})
provide a much better agreement with NLO DGLAP in the region of
large $x$.

We notice that our approach  is qualitatively close to the one
suggested in Ref.~\cite{kwec}, however they are not quite
identical: indeed they first differ in the treatment of $\alpha_s$
which is running in our picture in every Feynman graph involved,
whereas in Ref.~\cite{kwec} the DL contributions are obtained from
Refs.~\cite{ber} where $\alpha_s$ was kept fixed at an unknown
scale. To make it running, in Ref.~\cite{kwec} the parametrization
$\alpha_s = \alpha_s(Q^2)$ is suggested in the final formulae,
which is of course incorrect at small $x$ (see Refs.~\cite{egt1}
for details). In addition, the single- logarithmic contributions
were not accounted in Ref.~\cite{kwec} at all.

\section{Discussion and conclusions}
As is well known, the conventional DGLAP approach was originally
suggested for the region of rather large values of $x$, where
logarithms of $Q^2$ gave the most important contributions. At the
same time, DGLAP neglects the total resummation of logs of $x$. In
our approach we have been able to account for those contributions
to the non-singlet structure functions at small $x$. The results
are presented in Eqs.~(\ref{fnsint},\ref{gnsint}). Similarly to
DGLAP, the new anomalous dimensions (\ref{hns}) control the $Q^2$
-evolution while new coefficient functions (\ref{cns}) evolve the
initial quark densities from their empirical values $\delta q$ at
$x \sim 1$ to $x \ll 1$.

The extrapolation of Eqs.~(\ref{fnsint},\ref{gnsint}) into the
asymptotic region $x \to 0$  leads to the Regge asymptotic
formulae of Eqs.~(\ref{as},\ref{asympt}). The extrapolation of the
DGLAP expressions with the standard assumption of Eq.~(\ref{fita})
for the initial quark distribution $\delta q$ also leads to the
Regge behavior Eq.~(\ref{asdglap}). However, our approach provides
the Regge behavior because of the total resummation of the leading
logarithms of $x$ whereas the DGLAP Regge asymptotics is generated
by the phenomenological factor $x^{- \alpha}$ in the fit of
Eq.~(\ref{fita}) for $\delta q$. This allows us to conclude that
the singular part of the DGLAP input for the initial quark
densities mimics the impact of total resummation and can be
dropped when the resummation is done, simplifying the structure of
the input densities to expressions regular in $x$.
 It also turns out that the DGLAP small-$x$ asymptotics, using the fit
of Eq.~(\ref{fita}) for $\delta q$ for the non-singlet, is more
singular than our predictions. This contradicts the various
analyses of the experimental data of Refs.~\cite{kat}, which agree
with  our predictions. Our asymptotics also differ from DGLAP as
far as the $Q^2$ -dependence is concerned: indeed we predict an
asymptotic scaling, where the non-singlet structure functions
asymptotically depend on one argument $\xi = Q^2/x^2$ instead of
two arguments $Q^2$ and $x$, separately.

Giving up the singular factor from the DGLAP fit Eq.(\ref{fita})
and using its regular part $\phi(x)$ only, we have compared
numerically our formulae Eqs.~(\ref{fnsint},\ref{gnsint}) with the
NLO DGLAP results. Such a comparison confirms that in absence of
the singular terms the DGLAP small $x$-dependence and
$Q^2$-dependence of the non-singlets are slower than ours. The
results have been displayed in Figs.~1,2.

Finally, in order to obtain an overall approach which is consistent and
accurate in both regions
of small and large $x$, we have suggested to combine our results
and those of DGLAP by an appropriate definition of the anomalous
dimensions and the coefficient functions. Indeed, the expressions
of Eqs.~(\ref{newch}) for the improved anomalous dimensions and
coefficient functions account for both NLO DGLAP terms and our
total resummation of the most important logs of $x$. Therefore,
Eqs.~(\ref{fnsint},\ref{gnsint}) can be used both at large and
small $x$. Then the new inputs for $\delta q$ in this
approach could be much simpler functions of $x$ than in the standard
DGLAP approach.

\section{Acknowledgement}
The work is supported partly by grant RSGSS-1124.2003.2

\section{Figure captions}

Fig.~1: The $x$ -dependence  for: $R_1$ of Eq.~(\ref{r1})
(curve~1), $R_2$ of Eq.~(\ref{r2}) (curve~2) and $R_3$ of
Eq.~(\ref{r3}) (curve~3). All curves correspond to $Q^2 =
10$~GeV$^2$.

Fig.~2: The $Q^2$ -dependence for $R(Q^2)$ of Eq.~(\ref{rq}).
Curves 1,2,3 correspond to $x = 10^{-2},~10^{-3},~10^{-4}$
respectively .

\begin{figure}[p]
\epsfbox{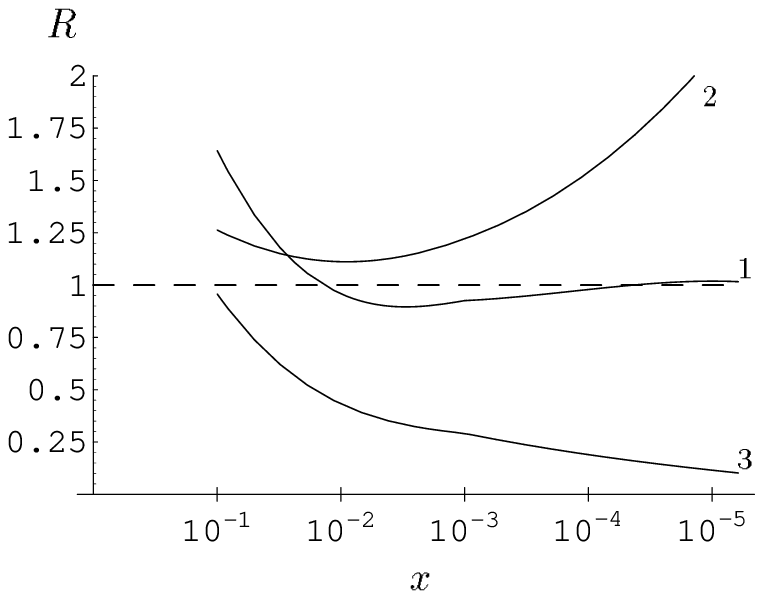} \caption{}
\end{figure}

\begin{figure}
\epsfbox{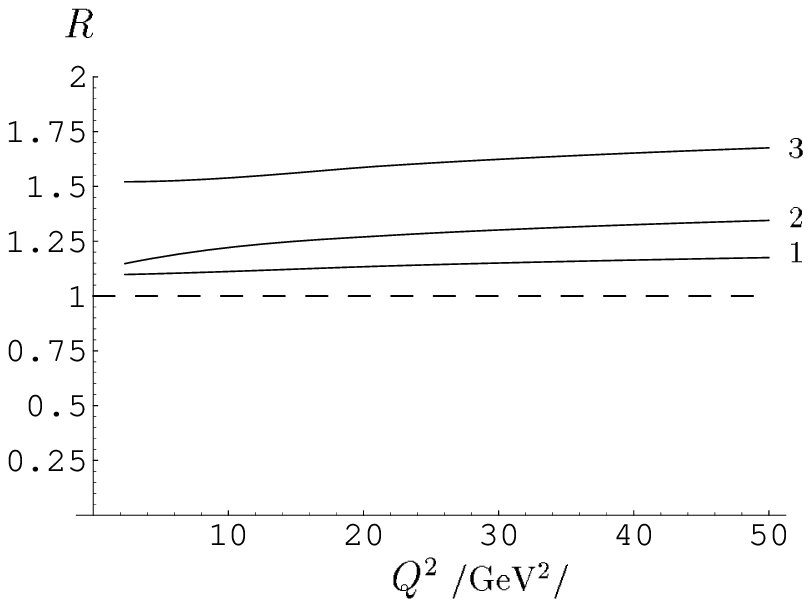} \caption{}
\end{figure}

\end{document}